\documentclass[10pt,conference]{IEEEtran}
\usepackage{epsfig,rotating,setspace,latexsym,amsmath,epsf,amssymb,amsfonts,bm,theorem,cite,caption,subcaption,enumerate,longtable,accents}
\usepackage{algorithm,algorithmic,graphicx,epsf,authblk,epstopdf,url,color,multirow,longtable}
\usepackage{mathtools}

\newtheorem{theorem}{Theorem}

\newtheorem{corollary}{Corollary}

\newtheorem{remark}{Remark}
\newtheorem{lemma}{Lemma}
\newtheorem{example}{Example}
\newenvironment{Proof}[1]{\medskip\par\noindent{\bf Proof:\,}\,#1}{{\mbox{\,$\blacksquare$}\par}}

\IEEEoverridecommandlockouts
\allowdisplaybreaks

\title{Symmetric Private Information Retrieval with User-Side Common Randomness}

\author{Zhusheng Wang \qquad Sennur Ulukus\\
	\normalsize Department of Electrical and Computer Engineering\\
	\normalsize University of Maryland, College Park, MD 20742\\
	\normalsize  \emph{zhusheng@umd.edu} \qquad \emph{ulukus@umd.edu}}

\begin{document}
\date{}
\maketitle

\begin{abstract}
We consider the problem of symmetric private information retrieval (SPIR) with user-side common randomness. In SPIR, a user retrieves a message out of $K$ messages from $N$ non-colluding and replicated databases in such a way that no single database knows the retrieved message index (user privacy), and the user gets to know nothing further than the retrieved message (database privacy). SPIR has a capacity smaller than the PIR capacity which requires only user privacy, is infeasible in the case of a single database, and requires shared common randomness among the databases. We introduce a new variant of SPIR where the user is provided with a random subset of the shared database common randomness, which is unknown to the databases. We determine the exact capacity region of the triple $(d, \rho_S, \rho_U)$, where $d$ is the download cost, $\rho_S$ is the amount of shared database (server) common randomness, and $\rho_U$ is the amount of available user-side common randomness. We show that with a suitable amount of $\rho_U$, this new SPIR achieves the capacity of conventional PIR. As a corollary, single-database SPIR becomes feasible. Further, the presence of user-side $\rho_U$ reduces the amount of required server-side $\rho_S$.  
\end{abstract}

\section{Introduction}
Private information retrieval (PIR) is a fundamental problem where a user downloads a message out of $K$ possible messages stored in $N$ non-colluding and replicated databases in such a way that no single database can know which message the user has downloaded \cite{PIR_ORI}. This privacy requirement is referred to as \emph{user privacy}. Symmetric PIR (SPIR) is an extended version of PIR where while downloading its desired message the user learns nothing about the remaining messages stored in the databases \cite{SPIR_ORI}. This is referred to as \emph{database privacy}. While PIR can be achieved with no shared common randomness, it is well-known that information-theoretic SPIR is possible only when the databases share a certain minimum amount of common randomness that is unknown to the user.

The information-theoretic capacity of PIR and SPIR have been found in \cite{PIR, SPIR} as $C_{\text{PIR}}=\frac{1-\frac{1}{N}}{1-\left(\frac{1}{N}\right)^K}$ and $C_{\text{SPIR}}=1-\frac{1}{N}$. First, $C_{\text{SPIR}}$ is smaller than $C_{\text{PIR}}$, since SPIR is a more constrained problem than PIR, as it requires not only user privacy but also database privacy. Second, single-database SPIR is infeasible as $C_{\text{SPIR}}=0$ for $N=1$, while single database PIR is feasible as $C_{\text{PIR}}=\frac{1}{K}$ for $N=1$. Our goal in this paper is two-fold: To explore ways to increase SPIR capacity to the level of PIR capacity, and perhaps more importantly, to make single-database SPIR feasible. 

Our motivation to focus on SPIR comes from constantly growing importance of privacy, not only the privacy of the retrieving user, but also the privacy of the databases, as the stored information in the databases may belong to other users. In addition, recent papers showed that other important privacy primitives, such as private set intersection (PSI) \cite{PSI_journal, MP-PSI_journal}, can be recast as versions of the SPIR problem, e.g., multi-message SPIR. Thus, here, we investigate ways to increase the SPIR capacity. Further, in practical applications, enforcing non-collusion could be difficult, as in some cases, all databases may naturally belong to the same entity, e.g., in various multi-party secure computation problems \cite{PSI_journal,MP-PSI_journal,FKN94,Secure_MPC,Secure_MPC_Review,FNP04,KS05,expand_Sun}. If all databases collude or belong to the same entity, the system essentially becomes a single-database system \cite{ColludingPIR}. The single-database PIR problem has been studied under extended conditions, e.g., side-information \cite{SDB_PIR, SDB_PIR_LRC, SDB_MMPIR1, SDB_MMPIR2}. Here, we investigate single-database SPIR under extended conditions, with the goal of making it feasible. Other important variations of PIR and SPIR problem have also been investigated; see e.g., \cite{one_extra_bit,SemanticPIR,PIR_coded,Kumar_PIRarbCoded,ChaoTian_coded_minsize,SPIR_coded,SPIR_Mismatched,MMPIR,MMPIR_PrivateSideInfo,BPIRjournal,SPIR_Eavesdropper,SPIR_Collusion,tandon_cache_2017,Cache-aided_PIR,PrefetchingPIR,PIR_cache_edge,PartialPSI_PIR,StorageConstrainedPIR_Wei,ChaoTian_leakage,leakyPIR,PIR_WTC_II,securestoragePIR,PIR_decentralized,HeteroPIR,Tian_upload,upload_SPIR,efficient_storage_ITW2019,StorageCost,Karim_nonreplicated,Tamo_journal,AsymmetryHurtsPIR,PrivateComputation,NoisyPIR,PrivateSearch}.

In this paper, we introduce SPIR with user-side common randomness, which solves the above two issues. In this model, the user obtains a part of the common randomness shared by the databases. The databases know the size of the user-side common randomness, but they do not know what the user possesses exactly. One way to implement this is for the user to fetch a part of the common randomness from the databases uniformly randomly, i.e., without the user knowing what it will get and without the databases knowing what it got, except for its cardinality. That is, all subsets of a certain size are equally likely to be obtained by the user. Another practical implementation could be for an external helper to distribute common randomness to the user and the databases randomly. 

For database-side (server-side) common randomness of amount $\rho_S$ and user-side common randomness of amount $\rho_U$, we determine the exact capacity region of the triple $(d, \rho_S, \rho_U)$, where $d$ is the download cost which is the inverse of the capacity. We show that with a suitable $\rho_U$, SPIR capacity becomes equal to the conventional PIR capacity. For the single-database case, since the conventional PIR capacity is $\frac{1}{K}$, this implies that single-database SPIR with user-side common randomness is feasible. In addition, the presence of user-side $\rho_U$ reduces the amount of required server-side $\rho_S$. 

\section{Problem Formulation}
We consider a system of $N \geq 1$ non-colluding databases each storing the same set of $K \geq 2$ i.i.d.~messages each of which consisting of $L$  i.i.d.~symbols uniformly selected from a sufficiently large finite field $\mathbb{F}_q$, i.e.,
\begin{align}
    H(W_k) &= L, \quad k \in [K] \label{Message Length} \\
    H(W_{1:K}) &= H(W_1) + \dots + H(W_K)  = KL \label{Message IID}
\end{align}

As in \cite{SPIR}, we use a random variable $\mathcal{F}$ to denote the randomness in the retrieval strategy implemented by the user. Due to the user privacy constraint, the realization of $\mathcal{F}$ is only known to the user, and is unknown to any of the databases. Due to the database privacy constraint, databases need to share some amount of common randomness $\mathcal{R}_S$; we will call this \emph{server-side} common randomness. Before the retrieval process starts, the user obtains a partial knowledge of $\mathcal{R}_S$. We denote it by $\mathcal{R}_U$, and call it \emph{user-side} common randomness. $\mathcal{R}_U$ is a subset of $\mathcal{R}_S$. The user-side common randomness $\mathcal{R}_U$ is unknown to the databases, i.e., it is only known to be equally distributed among all subsets of $\mathcal{R}_S$ with cardinality $|\mathcal{R}_U|$.

The message set $W_{1:K}$ stored in the databases is independent of the desired message index $k$, retrieval strategy randomness at the user $\mathcal{F}$ and all the common randomness, 
\begin{align} \label{Message Set Independence}
    I(W_{1:K};k,\mathcal{F},\mathcal{R}_S,\mathcal{R}_U) = 0, \quad \forall k, ~ \forall \mathcal{R}_U
\end{align}

During the query generation stage, the user has no access to the message set $W_{1:K}$ in the databases and the common randomness difference $\mathcal{R}_S \backslash \mathcal{R}_U$, 
\begin{align} \label{Query Independence}
    I(Q_{1:N}^{[k,\mathcal{R}_U]};W_{1:K},\mathcal{R}_S \backslash \mathcal{R}_U) = 0, \quad \forall k, ~ \forall \mathcal{R}_U
\end{align}

Using the desired message index and the user-side common randomness, the user generates a query for each database according to the retrieval strategy randomness $\mathcal{F}$. Hence, the queries $Q_n^{[k,\mathcal{R}_U]}, n \in [N]$ are deterministic functions of $\mathcal{F}$, 
\begin{align} \label{Deterministic Query}
    H(Q_1^{[k,\mathcal{R}_U]},Q_2^{[k,\mathcal{R}_U]},\dots,Q_N^{[k,\mathcal{R}_U]}|\mathcal{F}) = 0 \quad \forall k, ~ \forall \mathcal{R}_U
\end{align}

After receiving a query from the user, each database generates a truthful answer based on the stored message set and the server-side common randomness, 
\begin{align} \label{Deterministic Answer} 
    H(A_n^{[k,\mathcal{R}_U]}|Q_n^{[k,\mathcal{R}_U]},W_{1:K},\mathcal{R}_S) = 0, \quad \forall n, ~ \forall k, ~ \forall \mathcal{R}_U
\end{align}

After collecting all $N$ answers from the databases, the user should be able to decode the desired messages $W_{k}$ reliably, 
\begin{align} \label{Reliability} 
  \text{[reliability]} \quad &H(W_k|\mathcal{F},A_{1:N}^{[k,\mathcal{R}_U]},\mathcal{R}_U) = 0,  \quad \forall k, ~ \forall \mathcal{R}_U 
\end{align}

Due to the user privacy constraint, the query generated to retrieve the desired message should be statistically indistinguishable from other queries. Specifically, for all $k,k^\prime$, all $n$, and all user-side common randomness $\mathcal{R}_U$, there exists some $\mathcal{R}_U^\prime$ with $H(\mathcal{R}_U^\prime) = H(\mathcal{R}_U)$ such that,
\begin{align} \label{User Privacy}
    \text{[user privacy]} \quad  (&Q_n^{[k,\mathcal{R}_U]},A_n^{[k,\mathcal{R}_U]},W_{1:K},\mathcal{R}_S) \notag \\
    &\sim ~ (Q_n^{[k^\prime,\mathcal{R}_U^\prime]},A_n^{[k^\prime,\mathcal{R}_U^\prime]},W_{1:K},\mathcal{R}_S) 
\end{align}

Furthermore as in \cite{StorageCost}, after factorizing the joint distribution of all the random variables in the server, we obtain the following equivalent expression for user privacy for all potential query realizations $q$,
\begin{align}  \label{User Privacy 3}
    \text{[user privacy]} \quad P(Q_n^{[k,\mathcal{R}_U]}=q) = P(Q_n^{[k^\prime,\mathcal{R}_U^\prime]}=q)
\end{align}

Due to the database privacy constraint, the user should learn nothing about  $W_{\bar{k}}$ which is the complement of $W_{k}$, i.e., $W_{\bar{k}} = \{W_1,\cdots,W_{k-1},W_{k+1},\cdots,W_K\}$, 
\begin{align}
    \text{[database privacy]} \quad I(W_{\bar{k}};\mathcal{F},A_{1:N}^{[k,\mathcal{R}_U]},\mathcal{R}_U) = 0 \label{Server Privacy} 
\end{align}

Again due to the database privacy, the user should not gain any knowledge about the remaining common randomness in the server even after retrieving the desired message, 
\begin{align}
    I(\mathcal{R}_S \backslash \mathcal{R}_U;\mathcal{F},A_{1:N}^{[k,\mathcal{R}_U]},W_k,\mathcal{R}_U) = 0 \label{Common Randomness Difference Independence}
\end{align}

An achievable SPIR scheme is a scheme that satisfies the reliability constraint \eqref{Reliability}, the user privacy constraint \eqref{User Privacy} and the database privacy constraint \eqref{Server Privacy}. As usual, the efficiency of the scheme is measured in terms of the maximal number of downloaded bits by the user from all the databases, denoted by $D$. We define the normalized download cost $d$, the normalized server-side common randomness $\rho_S$, and the normalized user-side common randomness as $\rho_U$ as, 
\begin{align} \label{ratedefinition}
d = \frac{D}{L}, \quad \rho_S= \frac{H(\mathcal{R}_S)}{L}, \quad 
\rho_U= \frac{H(\mathcal{R}_U)}{L}
\end{align} 
where $L$ is the message length. Our goal in this paper is to determine the largest region for simultaneously achievable triples $(d, \rho_S, \rho_U)$ over all valid retrieval schemes.

\section{Main Results}
We state the main result of our paper in the following theorem which is the \emph{capacity region} for the triple $(d, \rho_S, \rho_U)$.

\begin{theorem} \label{theorem 1}
With user-side common randomness, the multi-database SPIR capacity region for $N \geq 2$ and $K \geq 2$ is 
\begin{align}
    & d \geq 1+\frac{1}{N}+\frac{1}{N^2}+\dots+\frac{1}{N^{K-1}} \label{thm1-1}\\
    & \rho_S - \rho_U \geq \frac{1}{N}+\frac{1}{N^2}+\dots+\frac{1}{N^{K-1}} \label{thm1-2} \\
    & \frac{N-1}{N}d +  \rho_U \geq 1 \label{thm1-3} \\
    & \frac{N}{N-1}\rho_U + N\rho_S \geq \frac{N}{N-1} \label{thm1-4}
\end{align}
\end{theorem}

\begin{remark}
The right hand side of (\ref{thm1-1}) is the optimum normalized download cost of classical PIR, $d_{\text{PIR}}$ \cite{PIR}. Thus, (\ref{thm1-1}) states that $d\geq d_{\text{PIR}}$. When $\rho_U=0$, i.e., when there is no user-side common randomness, (\ref{thm1-3}) becomes $d\geq d_{\text{SPIR}}$, where $d_{\text{SPIR}}=\frac{N}{N-1}$ is the optimum normalized download cost of classical SPIR \cite{SPIR}. Note that $d_{\text{SPIR}} > d_{\text{PIR}}$ for all $N$. Therefore, when $\rho_U=0$, (\ref{thm1-3}) is binding, (\ref{thm1-1}) is loose, and we have $d=d_{\text{SPIR}}$. Theorem~\ref{theorem 1} implies that with appropriate $\rho_U$, e.g., with $\rho_U=\frac{1}{N^K}$, both (\ref{thm1-1}) and (\ref{thm1-3}) can be made binding, at which time the new SPIR download cost achieves $d=d_{\text{PIR}}$. 
\end{remark}

\begin{remark}
When $\rho_U=0$, Theorem~\ref{theorem 1} reduces to the capacity of classical SPIR  \cite{SPIR}, as in this case, (\ref{thm1-3}) gives $d\geq \frac{N}{N-1}$, (\ref{thm1-4}) gives $\rho_S\geq \frac{1}{N-1}$, and  (\ref{thm1-1}) and (\ref{thm1-2}) are non-binding.
\end{remark}

\begin{remark}
The gap between $\rho_S$ and $\rho_U$ must be no smaller than a specific value as a function of $N$ and $K$ as given on the right hand side of (\ref{thm1-2}). This comes from the server privacy constraint, where part of the common randomness, i.e., $\mathcal{R}_S \backslash \mathcal{R}_U$, is utilized to hide the undesired messages.  
\end{remark}

\begin{remark}
From (\ref{thm1-4}), we observe that the existence of user-side common randomness can help reduce the required amount of server-side common randomness. For instance, for $N=2$ databases and $K=2$ messages, classical SPIR optimum download cost $d=d_{\text{SPIR}}=2$ is achieved by $\rho_S=1$ \cite{SPIR}. In Theorem~\ref{theorem 1}, $d=2$ can be achieved by $\rho_S=\frac{3}{4}$ with $\rho_U=\frac{1}{4}$.
\end{remark}

\begin{corollary} \label{corollary 1}
With user-side common randomness, the single-database SPIR capacity region for $N = 1$ and $K \geq 2$ is
\begin{align}
    & d \geq K \\
    & \rho_S - \rho_U \geq K-1 \label{cor1-2}\\
    & \rho_U \geq 1 
\end{align}
\end{corollary}

\begin{remark}
It is well-known that, for $N=1$, classical SPIR is not feasible \cite{SPIR}. With user-side common randomness, single-database SPIR becomes feasible.
\end{remark}

\begin{remark}
The optimal normalized download cost for single-database PIR is $d=K$ \cite{PIR,SDB_PIR}, which is achieved by downloading all messages from the server. One of the difficulties of single-database SPIR is that downloading all messages is not a valid SPIR scheme. Corollary~\ref{corollary 1} shows
that single-database PIR capacity can be achieved for single-database SPIR by means of user-side common randomness.
\end{remark}

\begin{remark}
The first two terms in Corollary~\ref{corollary 1} follow from the first two terms in Theorem~\ref{theorem 1}. The third term in Corollary~\ref{corollary 1} follows from the last two terms in Theorem~\ref{theorem 1} by multiplying both sides of the fourth term in Theorem~\ref{theorem 1} by $N-1$.
\end{remark} 

\begin{remark}
Like multi-database SPIR, in the single-database SPIR as well, the gap between $\rho_S$ and $\rho_U$ must be no smaller than a specific value as a function of $K$ as given in (\ref{cor1-2}) to avoid information leakage on undesired messages. 
\end{remark}

\section{Motivating Example}
\begin{example} \label{example 1}
We consider a single-database case $N = 1$, $K = 3$ and $L = 1$. We use $W_1$, $W_2$ and $W_3$ to denote the three messages. Our new achievable scheme is given in Table~\ref{table1}.

\begin{table}[h]
\begin{center}
\begin{tabular}{ |c|c|c|c| }
 \hline
 \multirow{2}{1.5em}{$\mathcal{R}_U$} & \multicolumn{3}{|c|}{desired message} \\
 \cline{2-4}
 & $W_1$ & $W_2$ & $W_3$ \\
 \hline
 $S_1$ & $W_1+S_1$ & $W_2+S_1$ & $W_3+S_1$ \\
 & $W_2+S_2$ & $W_3+S_2$ & $W_1+S_2$ \\
 & $W_3+S_3$ & $W_1+S_3$ & $W_2+S_3$ \\
 \hline
 $S_2$ & $W_1+S_2$ & $W_2+S_2$ & $W_3+S_2$ \\
 & $W_2+S_3$ & $W_3+S_3$ & $W_1+S_3$ \\
 & $W_3+S_1$ & $W_1+S_1$ & $W_2+S_1$ \\
 \hline
 $S_3$ & $W_1+S_3$ & $W_2+S_3$ & $W_3+S_3$ \\
 & $W_2+S_1$ & $W_3+S_1$ & $W_1+S_1$ \\
 & $W_3+S_2$ & $W_1+S_2$ & $W_2+S_2$ \\
 \hline
\end{tabular}
\end{center}
\vspace*{-0.1cm}
\caption{The query table for the case $N = 1$, $K = 3$.}
\label{table1}
\vspace*{-0.5cm}
\end{table}

The reliability constraint follows from the fact that the user can always decode the desired message by using its own common randomness. The server privacy constraint follows from the fact that the undesired messages are always mixed with unknown common randomness. For the user-privacy constraint, we have for all $k, k^\prime \in [3], k \neq k^\prime$ and a randomly selected $\mathcal{R}_U \in \{S_1,S_2,S_3\}$ under a uniform distribution, there exists another different $\mathcal{R}_U^\prime \in \{S_1,S_2,S_3\}$ such that,
\begin{align} 
    P(Q^{[k,\mathcal{R}_U]}=q) = P(Q^{[k^\prime,\mathcal{R}_U^\prime]}=q) = \frac{1}{3}
\end{align}
where $q \in \{[W_1+S_1,W_2+S_2,W_3+S_3],[W_1+S_2,W_2+S_3,W_3+S_1],[W_1+S_3,W_2+S_1,W_3+S_2]\}$. Specifically from the point of view of the server, the same set of queries can be invoked for any desired message $W_i, i \in [3]$ with the same probability distribution. This scheme achieves $d = 3$, $\rho_U = 1$ and $\rho_S = 3$, which exactly matches the boundary of the SPIR capacity region for $N = 1$ and $K = 3$ in Corollary~\ref{corollary 1}. 
\end{example}

\begin{example} \label{example 2}
We consider a multi-database case $N = 2$, $K = 2$ and $L = 4$. We use 
$[a_1, a_2, a_3, a_4]$ as a random uniform permutation of the symbols in 
the first message $W_1$, and independently, $[b_1, b_2, b_3, b_4]$ as another
one for $W_2$. Due to message index permutations, each set of queries represents one of $4! \cdot 4 \cdot 3 = 288$ different possible permutations. We have two different sets for each $\mathcal{R}_U$ because of necessary common randomness permutations. Our new achievable scheme for one random message index selection is given in Table~\ref{table2}.

\begin{table}[ht]
\begin{center}
\scalebox{0.88}{
\begin{tabular}{ |c|c|c|c|c| }
\hline
 \multirow{2}{1.5em}{$\mathcal{R}_U$} & \multicolumn{2}{|c|}{Desired message: $W_1$} & \multicolumn{2}{|c|}{Desired message: $W_2$}  \\
 \cline{2-5}
 & DB1 & DB2 & DB1 & DB2 \\
 \hline
 $S_1$ & $a_1+S_1$ & $a_2+S_1$  & $b_1+S_1$ & $b_2+S_1$  \\
 & $b_1+S_2$ & $b_2+S_3$  & $a_1+S_2$ & $a_2+S_3$  \\
 & $a_3+b_2+S_3$ & $a_4+b_1+S_2$  & $b_3+a_2+S_3$ & $b_4+a_1+S_2$  \\
 \cline{2-5}
 & $a_1+S_1$ & $a_2+S_1$  & $b_1+S_1$ & $b_2+S_1$  \\
 & $b_1+S_3$ & $b_2+S_2$  & $a_1+S_3$ & $a_2+S_2$  \\
 & $a_3+b_2+S_2$ & $a_4+b_1+S_3$  & $b_3+a_2+S_2$ & $b_4+a_1+S_3$  \\
 \hline
 $S_2$ & $a_1+S_2$ & $a_2+S_2$ & $b_1+S_2$ & $b_2+S_2$ \\
 & $b_1+S_3$ & $b_2+S_1$ & $a_1+S_3$ & $a_2+S_1$ \\
 & $a_3+b_2+S_1$ & $a_4+b_1+S_3$ & $b_3+a_2+S_1$ & $b_4+a_1+S_3$ \\
 \cline{2-5}
 & $a_1+S_2$ & $a_2+S_2$ & $b_1+S_2$ & $b_2+S_2$ \\
 & $b_1+S_1$ & $b_2+S_3$ & $a_1+S_1$ & $a_2+S_3$ \\
 & $a_3+b_2+S_3$ & $a_4+b_1+S_1$ & $b_3+a_2+S_3$ & $b_4+a_1+S_1$ \\
 \hline
 $S_3$ & $a_1+S_3$ & $a_2+S_3$ & $b_1+S_3$ & $b_2+S_3$  \\
 & $b_1+S_1$ & $b_2+S_2$ & $a_1+S_1$ & $a_2+S_2$ \\
 & $a_3+b_2+S_2$ & $a_4+b_1+S_1$ & $b_3+a_2+S_2$ & $b_4+a_1+S_1$ \\
 \cline{2-5} 
 & $a_1+S_3$ & $a_2+S_3$ & $b_1+S_3$ & $b_2+S_3$  \\
 & $b_1+S_2$ & $b_2+S_1$ & $a_1+S_2$ & $a_2+S_1$ \\
 & $a_3+b_2+S_1$ & $a_4+b_1+S_2$ & $b_3+a_2+S_1$ & $b_4+a_1+S_2$ \\
 \hline 
\end{tabular}
}
\end{center}
\vspace*{-0.1cm}
\caption{The query table for the case $N = 2$, $K = 2$.}
\label{table2}
\vspace*{-0.5cm}
\end{table}
\end{example}

Verification that this proposed scheme achieves the user privacy and the database privacy constraints is similar to the one in Example~\ref{example 1}. This scheme achieves $d = 1.5$, $\rho_U = 0.25$ and $\rho_S = 0.75$. This is a corner point of the capacity region in Theorem~\ref{theorem 1} where all inequalities are satisfied with equality. The other corner point when $\rho_U = 0$ is achieved by the classical SPIR scheme in \cite{SPIR}. Any point on the line segment joining these two points can be achieved by time-sharing between these two schemes. Any other remaining point in Theorem~\ref{theorem 1} can be achieved by adding extra randomness in the user- and server-side simultaneously, or by increasing the server-side common randomness and the download cost.

\section{Converse Proof}
We provide a sketch of the converse proof of Theorem~\ref{theorem 1} here. The four inequalities in Theorem~\ref{theorem 1} are proved in Lemmas~\ref{lemma 3}, \ref{lemma 4}, \ref{lemma 9} and \ref{lemma 10} below. Towards proving these four lemmas, we need Lemmas~\ref{lemma 1}-\ref{lemma 2} and Lemmas~\ref{lemma 5}-\ref{lemma 8} below. We note that Lemmas~\ref{lemma 1}-\ref{lemma 2} extend \cite[Lemmas~5-6]{PIR}, and Lemmas~\ref{lemma 5}-\ref{lemma 8} extend \cite[Eqns.~(26),~(27),~(30),~(39)]{SPIR}. These extensions are needed because we have two additional sets of random variables in our system model: $\mathcal{R}_S$ and  $\mathcal{R}_U$ with respect to techniques in \cite{PIR}, and $\mathcal{R}_U$ with respect to techniques in \cite{SPIR}.  

\begin{lemma} \label{lemma 1}
\begin{align}
    I(W_{2:K};Q_{1:N}^{[1,\mathcal{R}_U]},A_{1:N}^{[1,\mathcal{R}_U]},\mathcal{R}_S|W_1) \leq D - L
\end{align}
\end{lemma}

\begin{lemma} \label{lemma 2}
\begin{align}
    & I(W_{k:K};Q_{1:N}^{[k-1,\mathcal{R}_U]},A_{1:N}^{[k-1,\mathcal{R}_U]},\mathcal{R}_S|W_{1:k-1}) \notag \\
    & \quad \geq \frac{1}{N}I(W_{k+1:K};Q_{1:N}^{[k,\mathcal{R}_U^\prime]},A_{1:N}^{[k,\mathcal{R}_U^\prime]},\mathcal{R}_S|W_{1:k}) + \frac{L}{N}
\end{align}
\end{lemma}

\begin{lemma} [Minimal download cost $d$] \label{lemma 3}
\begin{align}
    d \geq 1+\frac{1}{N}+\frac{1}{N^2}+\dots+\frac{1}{N^{K-1}}
\end{align}
\end{lemma}
\begin{Proof}
Following steps similar to \cite[Eqns.~(62)-(67)]{PIR} for Lemma~\ref{lemma 2}, we obtain
\begin{align}
    & I(W_{2:K};Q_{1:N}^{[1,\mathcal{R}_U]},A_{1:N}^{[1,\mathcal{R}_U]},\mathcal{R}_S|W_1) \notag \\
    & \qquad \geq \left(\frac{1}{N}+\frac{1}{N^2}+\dots+\frac{1}{N^{K-1}}
    \right)L \label{lemma proof 2.1}
\end{align}
Combining the upper bound in Lemma~\ref{lemma 1} and the lower bound  in \eqref{lemma proof 2.1} completes the proof.
\end{Proof}

\begin{lemma} [Minimal difference between $\rho_S$ and $\rho_U$] \label{lemma 4}
\begin{align}
    \rho_S - \rho_U \geq \frac{1}{N}+\frac{1}{N^2}+\dots+\frac{1}{N^{K-1}}
\end{align}
\end{lemma}
\begin{Proof}
From \eqref{lemma proof 2.1}, we have the following relation,
\begin{align}
    & H(W_{2:K}|Q_{1:N}^{[1,\mathcal{R}_U]},A_{1:N}^{[1,\mathcal{R}_U]},W_1,\mathcal{R}_S) \notag \\
    & \quad \leq (K-1)L - \left(\frac{1}{N}+\frac{1}{N^2}+\dots+\frac{1}{N^{K-1}}\right)L \label{lemma proof 4.1}
\end{align}
Next, we have the following upper bound,
\begin{align}
    &I(W_{2:K};\mathcal{R}_S \backslash \mathcal{R}_U|Q_{1:N}^{[1,\mathcal{R}_U]},A_{1:N}^{[1,\mathcal{R}_U]},W_1,\mathcal{R}_U) \nonumber \\
    & \quad \leq H(\mathcal{R}_S \backslash \mathcal{R}_U|Q_{1:N}^{[1,\mathcal{R}_U]},A_{1:N}^{[1,\mathcal{R}_U]},W_1,\mathcal{R}_U)\\
    & \quad \stackrel{(\ref{Common Randomness Difference Independence})}{=} H(\mathcal{R}_S \backslash \mathcal{R}_U) 
    = H(\mathcal{R}_S) - H(\mathcal{R}_U) \label{lemma proof 4.2-1}
\end{align}
and the following lower bound,
\begin{align}
    &I(W_{2:K};\mathcal{R}_S \backslash \mathcal{R}_U|Q_{1:N}^{[1,\mathcal{R}_U]},A_{1:N}^{[1,\mathcal{R}_U]},W_1,\mathcal{R}_U) \nonumber \\
    & \stackrel{(\ref{Server Privacy}),(\ref{Reliability}),(\ref{Deterministic Query})}{=} (K-1)L - H(W_{2:K}|Q_{1:N}^{[1,\mathcal{R}_U]},A_{1:N}^{[1,\mathcal{R}_U]},W_1,\mathcal{R}_S) \label{lemma proof 4.3}\\
    & \quad \stackrel{\eqref{lemma proof 4.1}}{\geq} \left(\frac{1}{N}+\frac{1}{N^2}+\dots+\frac{1}{N^{K-1}}\right)L \label{lemma proof 4.4}
\end{align}
Combining (\ref{lemma proof 4.2-1}) and (\ref{lemma proof 4.4}) yields the desired result.
\end{Proof}

\begin{lemma} \label{lemma 5} For any $k^\prime \neq k$,
\begin{align} 
    & H(A_n^{[k,\mathcal{R}_U]}|Q_n^{[k,\mathcal{R}_U]},W_k,\mathcal{R}_U) \notag \\
    & \quad \geq H(A_n^{[k^\prime,\mathcal{R}_U^\prime]}|Q_n^{[k^\prime,\mathcal{R}_U^\prime]},W_k,\mathcal{R}_U^\prime) - H(\mathcal{R}_U) \label{lemma eqn 5}
\end{align}
\end{lemma}

\begin{lemma} \label{lemma 6}
    \begin{align}
    H(A_n^{[k,\mathcal{R}_U]}|Q_n^{[k,\mathcal{R}_U]},\mathcal{R}_U)
    &= H(A_n^{[k^\prime,\mathcal{R}_U^\prime]}|Q_n^{[k^\prime,\mathcal{R}_U^\prime]},\mathcal{R}_U^\prime) \label{lemma eqn 6}
    \end{align}
\end{lemma}

\begin{lemma} \label{lemma 7}
 \begin{align}
   & H(A_n^{[k,\mathcal{R}_U]}|\mathcal{F},Q_n^{[k,\mathcal{R}_U]},W_k,\mathcal{R}_U) \notag \\
   & \quad  = H(A_n^{[k,\mathcal{R}_U]}|Q_n^{[k,\mathcal{R}_U]},W_k,\mathcal{R}_U) \label{lemma eqn 7}
\end{align}
\end{lemma}

\begin{lemma} \label{lemma 8} For any $k^\prime \neq k$,
   \begin{align}
    H(A_n^{[k^\prime,\mathcal{R}_U^\prime]}|Q_n^{[k^\prime,\mathcal{R}_U^\prime]},\mathcal{R}_U^\prime) = H(A_n^{[k^\prime,\mathcal{R}_U^\prime]}|Q_n^{[k^\prime,\mathcal{R}_U^\prime]},W_k,\mathcal{R}_U^\prime) \label{lemma eqn 8}
    \end{align}
\end{lemma}

\begin{lemma} [Minimal bound for $d$ and $\rho_U$] \label{lemma 9}
   \begin{align}
       \frac{N-1}{N}d +  \rho_U \geq 1   
   \end{align}
\end{lemma}
\begin{Proof}
Starting from the message length assumption \eqref{Message Length},
\begin{align}
    L & = H(W_k) \stackrel{(\ref{Message Set Independence})}{=} H(W_k|\mathcal{F},\mathcal{R}_U) \\
      & \stackrel{(\ref{Reliability})}{=} H(W_k|\mathcal{F},\mathcal{R}_U) - H(W_k|\mathcal{F},A_{1:N}^{[k,\mathcal{R}_U]},\mathcal{R}_U) \\
      &= I(W_k;A_{1:N}^{[k,\mathcal{R}_U]}|\mathcal{F},\mathcal{R}_U) \\
      &= H(A_{1:N}^{[k,\mathcal{R}_U]}|\mathcal{F},\mathcal{R}_U) - H( A_{1:N}^{[k,\mathcal{R}_U]}|\mathcal{F},W_k,\mathcal{R}_U) \\
      &\leq H(A_{1:N}^{[k,\mathcal{R}_U]}|\mathcal{F},\mathcal{R}_U) - H( A_n^{[k,\mathcal{R}_U]}|\mathcal{F},Q_n^{[k,\mathcal{R}_U]},W_k,\mathcal{R}_U) \\
      & \stackrel{(\ref{lemma eqn 7})}{=} H(A_{1:N}^{[k,\mathcal{R}_U]}|\mathcal{F},\mathcal{R}_U) - H( A_n^{[k,\mathcal{R}_U]}|Q_n^{[k,\mathcal{R}_U]},W_k,\mathcal{R}_U) \label{lemma proof 9.1} \\
      & \stackrel{(\ref{lemma eqn 5})}{\leq} H(A_{1:N}^{[k,\mathcal{R}_U]}|\mathcal{F},\mathcal{R}_U) -  H(A_n^{[k^\prime,\mathcal{R}_U^\prime]}|Q_n^{[k^\prime,\mathcal{R}_U^\prime]},W_k,\mathcal{R}_U^\prime) \notag \\
      & \quad + H(\mathcal{R}_U) \\
      & \stackrel{(\ref{lemma eqn 8})}{=} H(A_{1:N}^{[k,\mathcal{R}_U]}|\mathcal{F},\mathcal{R}_U) - H(A_n^{[k^\prime,\mathcal{R}_U^\prime]}|Q_n^{[k^\prime,\mathcal{R}_U^\prime]},\mathcal{R}_U^\prime) \notag \\
      & \quad + H(\mathcal{R}_U) \\
      & \stackrel{(\ref{lemma eqn 6})}{=} H(A_{1:N}^{[k,\mathcal{R}_U]}|\mathcal{F},\mathcal{R}_U) - H(A_n^{[k,\mathcal{R}_U]}|Q_n^{[k,\mathcal{R}_U]},\mathcal{R}_U) \notag \\
      & \quad + H(\mathcal{R}_U) \label{lemma proof 9.2} \\      
      & \stackrel{(\ref{Deterministic Query})}{\leq}  \!\! H(A_{1:N}^{[k,\mathcal{R}_U]}|\mathcal{F},\mathcal{R}_U) \!\!-\!\! H(A_n^{[k,\mathcal{R}_U]}|\mathcal{F},\mathcal{R}_U) \!\!+\!\! H(\mathcal{R}_U) \label{lemma proof 9.3} 
\end{align}
By summing \eqref{lemma proof 9.3} over all $n \in [1:N]$, we obtain the following relationship, which completes the proof,
\begin{align}
    NL &\leq NH(A_{1:N}^{[k,\mathcal{R}_U]}|\mathcal{F},\mathcal{R}_U) - \sum\limits_{n=1}^N H(A_n^{[k,\mathcal{R}_U]}|\mathcal{F},\mathcal{R}_U) \notag \\
    & \quad + NH(\mathcal{R}_U)  \\
    &\leq (N-1)H(A_{1:N}^{[k,\mathcal{R}_U]}|\mathcal{F},\mathcal{R}_U) + NH(\mathcal{R}_U) \label{lemma proof 9.4} \\
    &\leq (N-1)\sum\limits_{n=1}^N H(A_n^{[k,\mathcal{R}_U]}|\mathcal{F},\mathcal{R}_U) + NH(\mathcal{R}_U) \\
    &\leq (N-1)D +  NH(\mathcal{R}_U)
\end{align}
\end{Proof} 

\begin{lemma} [Minimal bound for $\rho_U$ and $\rho_S$] \label{lemma 10}
   \begin{align}
       \frac{N}{N-1}\rho_U + N\rho_S \geq \frac{N}{N-1}  
   \end{align}
\end{lemma}
\begin{Proof}
Starting with the database privacy constraint \eqref{Server Privacy},
\begin{align}
    0 &= I(W_{\bar{k}};\mathcal{F},A_{1:N}^{[k,\mathcal{R}_U]},\mathcal{R}_U) \\
      & \stackrel{(\ref{Message Set Independence})}{=} I(W_{\bar{k}};A_{1:N}^{[k,\mathcal{R}_U]},\mathcal{R}_U|\mathcal{F}) \\
      & \stackrel{(\ref{Reliability})}{=} I(W_{\bar{k}};A_{1:N}^{[k,\mathcal{R}_U]},\mathcal{R}_U|\mathcal{F}) +  I(W_{\bar{k}};W_k|\mathcal{F},A_{1:N}^{[k,\mathcal{R}_U]},\mathcal{R}_U) \\
      &= I(W_{\bar{k}};A_{1:N}^{[k,\mathcal{R}_U]},W_k,\mathcal{R}_U|\mathcal{F}) \\
      &= I(W_{\bar{k}};A_{1:N}^{[k,\mathcal{R}_U]}|\mathcal{F},W_k,\mathcal{R}_U) + I(W_{\bar{k}};W_k,\mathcal{R}_U|\mathcal{F}) \\
      & \stackrel{(\ref{Message Set Independence})}{=} I(W_{\bar{k}};A_{1:N}^{[k,\mathcal{R}_U]}|\mathcal{F},W_k,\mathcal{R}_U) \\
      & \geq I(W_{\bar{k}};A_n^{[k,\mathcal{R}_U]}|\mathcal{F},W_k,\mathcal{R}_U) \\
      & \stackrel{(\ref{Deterministic Answer}),(\ref{Deterministic Query})}{=} H(A_n^{[k,\mathcal{R}_U]}|\mathcal{F},W_k,\mathcal{R}_U) - H(A_n^{[k,\mathcal{R}_U]}|\mathcal{F},W_{1:K},\mathcal{R}_U) \notag \\
      & \quad + H(A_n^{[k,\mathcal{R}_U]}|\mathcal{F},W_{1:K},\mathcal{R}_S) \\
      & \geq H(A_n^{[k,\mathcal{R}_U]}|\mathcal{F},W_k,\mathcal{R}_U) - H(A_n^{[k,\mathcal{R}_U]}|\mathcal{F},W_{1:K},\mathcal{R}_U) \notag \\
      & \quad + H(A_n^{[k,\mathcal{R}_U]}|\mathcal{F},W_{1:K},\mathcal{R}_S,\mathcal{R}_U) \\
      &= H(A_n^{[k,\mathcal{R}_U]}|\mathcal{F},W_k,\mathcal{R}_U) - I(A_n^{[k,\mathcal{R}_U]};\mathcal{R}_S|\mathcal{F},W_{1:K},\mathcal{R}_U) \\
      &= H(A_n^{[k,\mathcal{R}_U]}|\mathcal{F},W_k,\mathcal{R}_U) - H(\mathcal{R}_S|\mathcal{F},W_{1:K},\mathcal{R}_U) \notag \\
      & \quad + H(\mathcal{R}_S|\mathcal{F},A_n^{[k,\mathcal{R}_U]},W_{1:K},\mathcal{R}_U) \\
      & \geq H(A_n^{[k,\mathcal{R}_U]}|\mathcal{F},W_k,\mathcal{R}_U) - H(\mathcal{R}_S|\mathcal{F},W_k,\mathcal{R}_U) \\
      &= H(A_n^{[k,\mathcal{R}_U]}|\mathcal{F},W_k,\mathcal{R}_U) - H(\mathcal{R}_U,\mathcal{R}_S \backslash \mathcal{R}_U|\mathcal{F},W_k,\mathcal{R}_U) \\
      &= H(A_n^{[k,\mathcal{R}_U]}|\mathcal{F},W_k,\mathcal{R}_U) - H(\mathcal{R}_S \backslash \mathcal{R}_U|\mathcal{F},W_k,\mathcal{R}_U) \\
      & \stackrel{(\ref{Common Randomness Difference Independence})}{=} H(A_n^{[k,\mathcal{R}_U]}|\mathcal{F},W_k,\mathcal{R}_U) - H(\mathcal{R}_S \backslash \mathcal{R}_U) \\
      &=  H(A_n^{[k,\mathcal{R}_U]}|Q_n^{[k,\mathcal{R}_U]},\mathcal{R}_U) - H(\mathcal{R}_S)  \label{lemma proof 10.1}
\end{align}
where \eqref{lemma proof 10.1} follows from the steps between \eqref{lemma proof 9.1}-\eqref{lemma proof 9.2} by applying Lemma~\ref{lemma 5} to Lemma~\ref{lemma 8} again.

By summing \eqref{lemma proof 10.1} over all $n \in [1:N]$, we obtain the following relationship, which completes the proof,
\begin{align}
    0 &\geq \sum\limits_{n=1}^N H(A_n^{[k,\mathcal{R}_U]}|Q_n^{[k,\mathcal{R}_U]},\mathcal{R}_U) - NH(\mathcal{R}_S) \\
      &\geq H(A_{1:N}^{[1,\mathcal{R}_U]}|\mathcal{F},Q_n^{[1,\mathcal{R}_U]},\mathcal{R}_U)  - NH(\mathcal{R}_S) \\
      & \stackrel{(\ref{Deterministic Query}),\eqref{lemma proof 9.4}}{\geq} \frac{N}{N-1}L - \frac{N}{N-1}H(\mathcal{R}_U) - NH(\mathcal{R}_S)
\end{align}
\end{Proof}

\section{Achievability Proof}
Our achievability is based on the principle of converting a given PIR scheme into a valid SPIR scheme using the server-side and user-side common randomness in a manner that does not compromise the download cost; e.g., \cite{OTPE}. To that goal, the common randomness added to the desired symbols are substracted out as they are available at the user side, and the remaining common randomness unknown to the user protects the undesired messages. The challenge is to implement this for all possible user-side common randomness realizations which are unknown ahead of time. Steps of our achievable scheme: 
\begin{enumerate}
    \item \emph{Initial PIR query generation:} For given $N$ and $K$, generate an initial PIR query table for each desired message using the scheme in \cite{PIR}, e.g., Tables~\ref{table1}-\ref{table2} without $S_i$s.
    \item \emph{Server-side common randomness assignment:} For each desired message and each permutation of message index (e.g., 288 permutations in Example~\ref{example 2}), mix all 1-sum symbols from the desired message across all the databases with the same new common randomness. We call it seed common randomness (e.g., $S_1$ in first three rows of Table~\ref{table2}). Assign a new distinct common randomness to every 1-sum symbol from the undesired messages. For every $k$-sum symbol containing a desired message symbol, mix it with the common randomness from the $(k\!-\!1)$-sum symbol having the same $k\!-\!1$ undesired message symbols queried at another database. For every $k$-sum symbol not containing any desired message symbol, assign a new distinct common randomness. Repeat this until $k$ reaches $K$. Next, we permute non-seed common randomness indices (e.g., $4$th-$6$th rows of Table~\ref{table2}). We call this whole modified query table a \emph{query cell}. 
    \item \emph{Server-side common randomness cycling:} While keeping each query cell,  create a new one by adding 1 (mod $|\mathcal{R}_S|$) to each common randomness index (e.g., $S_1$ becomes $S_2$ in Table~\ref{table2}). Repeat it $|\mathcal{R}_S|$ times such that each query cell has a different seed common randomness index.
    \item \emph{Query cell determination:} The user has $|\mathcal{R}_U|$ server-side common randomness. The user determines the query cell to be invoked, and selects a random permutation within that cell, by matching its user-side common randomness to the seed common randomness of the cell. 
\end{enumerate}

In this scheme, the message length $L$ is $N^K$ as in \cite{PIR}, the total amount of server-side common randomness required $|\mathcal{R}_S|$ is $1+\cdots+N^{K-1}$ and the total amount of user-side common randomness required $|\mathcal{R}_U|$ is $1$. 

\bibliographystyle{unsrt}
\bibliography{ISIT2021}

\end{document}